\documentclass[runningheads,a4paper]{llncs}
\usepackage[utf8x]{inputenc}
\usepackage[ruled,linesnumbered,vlined]{algorithm2e}

\usepackage{fancyhdr}
\usepackage[pdftex]{graphicx}
\usepackage{longtable}
\usepackage{mathtools}
\usepackage{amsmath}
\usepackage{pbox}
\usepackage{amsfonts}
\usepackage{subcaption}
\usepackage[pdftex,bookmarks=true,pdfborder={0 0 0}]{hyperref} %omogoca zaznamke 

\DeclareGraphicsExtensions{.pdf,.png,.jpg}
\graphicspath{{./images/}{./benchmarks/}}
\newtheorem{assumption}{Assumption}

\usepackage{url}
\urldef{\mailsa}\path|{matevz.jekovec,andrej.brodnik}fri.uni-lj.si|
\newcommand{\keywords}[1]{\par\addvspace\baselineskip
\noindent\keywordname\enspace\ignorespaces#1}

\begin{document}

\mainmatter

\title{ERA Revisited:\\Theoretical and Experimental Evaluation}
\titlerunning{ERA Revisited: Theoretical and Experimental Evaluation}

\author{Matevž Jekovec and Andrej Brodnik}
\institute{University of Ljubljana, Faculty of Computer and Information Science, Slovenia\\
\mailsa\\
%\mailsb
}

\toctitle{ERA Revisited: Theoretical and Experimental Evaluation}
\tocauthor{Matevž Jekovec and Andrej Brodnik}
\maketitle

\begin{abstract}

Efficient construction of the suffix tree given an input text is an active area of research from the time it was first introduced. Both theoretical computer scientists and engineers tackled the problem. In this paper we focus on the fastest practical suffix tree construction algorithm to date, ERA. We first provide a theoretical analysis of the algorithm assuming the uniformly random text as an input and using the PEM model of computation with respect to the lower bounds. Secondly, we empirically confirm the theoretical results in different test scenarios exposing the critical terms. Thirdly, we discuss the fundamental characteristics of the input text where the fastest suffix tree construction algorithms in practice fail. This paper serves as a foundation for further research in the parallel text indexing area.

\end{abstract}

\section{Introduction}

The suffix tree \cite{Weiner1973} and the suffix array \cite{Manber1990} data structures are the most widely used data structures in text indexing applications. They allow answering three main queries: 1) is the given query string $P$ present in the text, 2) where are all its' occurrences located in the text, and 3) finding the longest prefix of the given query string $P$ in the text. Both, suffix trees and suffix arrays with longest common prefix (LCP) information allow answering these questions in time $O(|P|)$ \cite{Abouelhoda2004}. Through the rest of the paper, we will focus on the suffix tree construction algorithms only. Note that all the lower bounds were shown to hold for the suffix arrays as well.

The theoretical lower bound for the suffix tree construction is inherently bounded by the integer sorting operation (see \cite{Farach-Colton2000}). For bounded alphabets, the optimal suffix tree construction algorithms requiring $\Theta(N)$ time were already designed back in 1970s by Weiner \cite{Weiner1973}. For unbounded alphabets, a comparison-based sorting algorithm needs to be employed thus the $\Omega(N \lg N)$ time and $\Omega(\frac{N}{B} \log_{M/B} \frac{N}{B})$ I/Os in DAM model of Aggarwal and Vitter \cite{Aggarwal1988} are required where $N$, $M$ and $B$ denote the input text length, cache size and block size respectively. Farach-Colton constructed the first time and I/O optimal suffix tree construction algorithm for unbounded alphabets. Even more, he also suggested the algorithm can run in $\Theta(\frac{N}{DB} \log_{M/B} \frac{N}{B})$ I/Os using $D$ disks in the Parallel DAM model of Vitter and Shriver \cite{Vitter1994}. Unfortunately, there is no practical implementation of his algorithm, so we have no idea whether the algorithm performs fast in practice.

Looking at the carefully engineered suffix tree construction algorithms used in practice for indexing the human genome and proteins, basically all of them require $O(N^2)$ execution time in the worst case, yet the worst case ``never occur in practice''. The most recent practical algorithms are Big string, Big Suffix Tree (B$^2$ST) \cite{Barsky2009}, Wavefront (WF) \cite{Ghoting2009}, Elastic Range (ERA) \cite{Mansour2011}, and Parallel Continuous Flow (PCF) \cite{Comin2013}. While the authors of PCF claim the algorithm is well scalable on supercomputers and large clusters, our algorithm of choice is ERA because of its better performance on multicore machines. Also, in the private conversation with Timo Bingmann, we compared ERA running on a single core to the fastest (sequential) suffix array and LCP construction algorithm to date, eSAIS \cite{Bingmann2013-2}. ERA was around 2-times faster using the human genome as an input. 

The goal of this paper is threefold: Firstly, the original paper on ERA only provided a very vague worst case time analysis for extremely skewed input texts. Intrigued by its obvious speed, we provide a thorough parallel time and I/O complexity analysis assuming the uniformly random input text and examine whether the result is close to the integer sorting lower bound. We picked the random input string because the typical use cases for ERA and similar text indexing algorithms are human genome, DNA, proteins and music which were experimentally shown in Heinz et al.\ paper \cite{Heinz2002} they were \emph{practically} random. Secondly, we perform a number of empirical tests confirming the theoretical results in different test scenarios exposing different critical terms, and also providing insight information of possible algorithm bottlenecks. Thirdly, we discuss the fundamental characteristics of the input text where, not only ERA, but all of the practical suffix tree construction algorithm implementations fail. Our long-pursuing goal is to design a theoretically optimal parallel suffix tree construction algorithm, which is also the fastest in practice. This paper gives the intuition how the practical algorithms should be designed with respect to the theoretical lower bounds.

The rest of the paper is divided into five sections. In \autoref{chap:background} we introduce our notation, the suffix tree, PEM model of computation and outline the ERA algorithm. In \autoref{chap:analysis} we provide theoretical time and I/O complexity analysis. In \autoref{chap:empirical_evaluation} we present the experimental evaluation followed by a discussion and a conclusion in \autoref{chap:discussion_and_conclusion}.

\section{Background}\label{chap:background}

\subsection{Suffix tree}

Given an input string $S[1..N]$, the substring $S[i..N]$ for any $i \in \lbrace 1 ... N \rbrace$ is called a suffix of $S$. All characters in $S$ are from a finite alphabet $\Sigma$ of size $\sigma$ except for the last character $S[N]=\$$ which is called the \emph{delimiter character} and is unique in the text. The \emph{suffix tree} (formally introduced and constructed in \cite{Weiner1973}) is a path compressed trie storing all suffixes of $S$. Each edge represents a substring of $S$ of length $1 ... N$. There are exactly $N$ leaves in the suffix tree where each leaf stores a position of the suffix in the original text. Each path from the root to the leaf defines a unique suffix and the value in the leaf determines its location inside the text $S$. The children of each node are lexicographically ordered. If we traverse all the leaves from left to right, we obtain lexicographically ordered suffixes and their positions in the text i.e. the values of the suffix array.

\subsection{PEM model}

\begin{figure}[htb]
  \begin{center}
    \leavevmode
    \includegraphics[width=9cm]{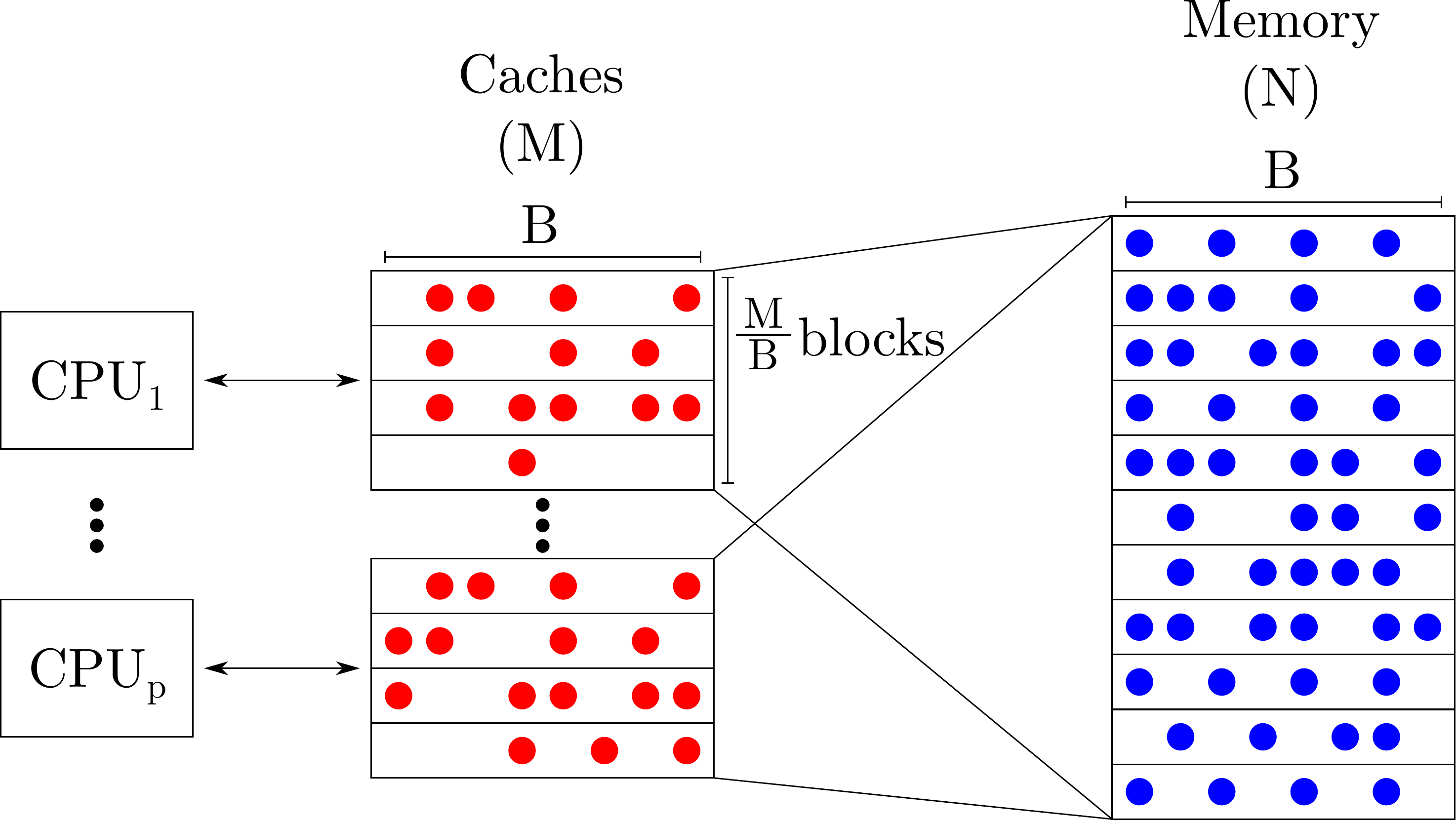}
  \end{center}
  \caption{The Parallel External Memory model.}
  \label{fig:pem}
\end{figure}

Parallel external memory (PEM) model \cite{Arge2008} is a version of a shared memory model of computation consisting of $p$ processors and a two-level memory hierarchy. The $2^{nd}$ level is an external memory and accessible by any processor, whereas the $1^{st}$ level memories are private caches, each of size $M$. Processors can only perform operations by accessing their caches. The data is transferred between the external memory and the caches in both directions in blocks of size $B$. It is assumed there is enough bandwidth between the external memory and caches for transferring any block to each of the processors in parallel. Figure \ref{fig:pem} illustrates the PEM model.

If concurrent writes to the external memory occur, any of the CRCW, CREW or EREW models are allowed. If not stated otherwise, in this paper we will use PEM CREW model. In our analysis we will use two performance measures: 1) the parallel execution time measuring the number of executed instructions in parallel, and 2) the parallel I/O complexity measuring the number of parallel block transfers between the external and the main memory.

\subsection{ERA}

Elastic Range algorithm (ERA), introduced by Mansour et al.\ in \cite{Mansour2011}, is currently the fastest practical algorithm for the suffix tree construction. It assumes both the input string and the resulting suffix tree are too big to fit into the main memory ($N>M$). It runs in two phases called the \emph{vertical partitioning} and the \emph{horizontal partitioning}.

The vertical partitioning is done sequentially. It is used to partition the suffix tree into manageable suffix subtrees $\mathcal{T}_{\pi}$, where $\mathcal{T}_{\pi}$ corresponds to a prefix $\pi$ and is small enough to fit into the main memory of size $M$. Let $f(\pi)$ denote the frequency of the prefix $\pi$ in $S$. To determine each $\mathcal{T}_{\pi}$ size without building the actual subtree, the algorithm uses the following assumption

\begin{assumption}
\label{assumption:subtree-size}
If each occurrence of the prefix $\pi$ in the text corresponds to a leaf in the resulting suffix tree $\mathcal{T}$, and the largest number of internal nodes is the number of leafs minus $1$, then the size of the suffix subtree $|\mathcal{T}_{\pi}| \approx 2 f(\pi)$, where $f(\pi)$ is the frequency of a prefix $\pi$ in the text.
\end{assumption}

In order to optimize the memory usage even further, the algorithm uses a first-fit bin packing heuristic to form \emph{virtual trees} of one or more $\mathcal{T}_{\pi}$ filling the main memory as tight as possible. The result of the vertical partitioning is the top part of the suffix tree (an uncompacted trie) stored to the disk and a working set of virtual trees.

The horizontal partitioning is done in parallel. Each processor takes one virtual tree from the working set and constructs the suffix subtree $\mathcal{T}_{\pi}$, one for each prefix inside the virtual tree. Note that each processor has access to the whole input string $S$. The basic essence of the efficient horizontal partitioning is optimizing 1) the input string access and 2) the main memory access where the suffix tree is being constructed. Goal 1) is achieved by constructing the suffix tree in a level-order manner and using scans only over the input text, one scan for each suffix tree level. Achieving goal 2) is more demanding. The suffix tree contains suffixes ordered lexicographically, whereas the original text can have any character present at arbitrary location. Therefore, one cannot expect to directly construct the suffix tree using a tree traversal and a single scan of the input string. ERA first reads the input string chunks using scans only, in-memory lexicographically sorts them and then outputs the suffix subtree with a single tree traversal. The main memory is thus used as a buffer where all the random accesses occur whereas the external memory is accessed sequentially only. Finally, gluing all the suffix subtrees together from all the virtual trees is done implicitly by storing each suffix subtree $\mathcal{T}_{\pi}$ to a separate file named $\pi$.

\section{Theoretical Analysis}
\label{chap:analysis}

For the analysis we acknowledge the following assumption
\begin{assumption}
\label{assumption:random-text}
The input text is a uniformly random input string such that the probability of each character to occur at any place is $\frac{1}{\sigma}$.
\end{assumption}

It was shown in \cite{Szpankowski1993} the suffix tree of uniformly random string of length $N$ is balanced with expected height $\lceil \log_\sigma N \rceil$. Suffixes of such input string also have the shortest longest common prefixes of expected length $\lceil \log_\sigma N \rceil - 2$. Because the lengths of the longest common prefixes determine the minimal work needed to discriminate between the suffixes in order to sort them, and because sorting is always at least implicitly used in text indexing, the random input text is indeed the \emph{best case input}. Nevertheless, it was shown in \cite[Table II]{Heinz2002} that for practical usage the random input string captures well the characteristics of a single instance of the human genome, DNA in general, proteins or western music in MIDI representation. We discuss situations of a more skewed input text in \autoref{chap:discussion_and_conclusion}.

The pointer-based implementation of a node in the suffix tree requires at least $2 \lg N$ bits since it holds one pointer to a child node or the input string, and a skip value. We assume the following precondition:

\begin{assumption}
\label{assumption:block-size}
The block of size $B$ must fit at least one node: $B \geq 2\lg N$
\end{assumption}

%\begin{assumption}
%The alphabet $\Sigma$ is of finite size $\sigma < M$. %$M \geq 2B$.
%\end{assumption}

%The size of the alphabet increases the search space significantly since the vertical partitioning generates and counts all possible S-prefixes of given length. This assumption is not mandatory, but greatly simplifies the final steps in the vertical partitioning time analysis.

\subsection{Vertical Partitioning}
\label{sec:vertical_partitioning}

Algorithm \ref{alg:verticalpartitioning} shows the vertical partitioning algorithm from \cite[pp.\ $4$]{Mansour2011} with a slightly different notation. First we analyse the external loop execution in lines $4$-$11$. The algorithm builds a set of prefixes $P=\lbrace \pi_1, \pi_2, ... \rbrace$ corresponding to suffix subtrees small enough to fit into the main memory. Taking the Assumption \ref{assumption:random-text} into account, the frequency of any $|\pi_i|=1$ in the first iteration is expected $f(\pi_i)=\frac{N}{\sigma}$. By extending $\pi_i$ for a single character, $\pi_i$ is replaced by $\sigma$ new prefixes with their frequency decreased for an expected factor $\sigma$ at each step. We decrease the prefix frequency until the following conditions are met: 1) the corresponding subtree under Assumption \ref{assumption:subtree-size} fits the main memory of size $M$, and 2) all the input string chunks of length $B$ for each occurrence of the prefix, fit $M$. We write this as $f(\pi_i) \leq \frac{M}{max(B, 2 \lg N)}$. Under Assumption \ref{assumption:block-size} follows $max(B, 2\lg N)=B$.

The external loop repeats until the expected subtree size becomes $f(\pi_i) \leq \frac{M}{B}$. This leads to the total expected number of iterations $\log_\sigma N - \log_\sigma M/B = \log_\sigma \frac{NB}{M}$ and obtaining $|P| = \frac{NB}{M}$ unique prefixes.

The inner loop in lines $7$-$10$ extends intermediate $\pi_i \in |P'|$. It is expected to repeat $|P'|=\sigma^j$ times every $j^{th}$ iteration of the external loop. Each internal iteration takes $\sigma$ time to extend the current prefix $\pi_i$ except for the cases when $f(\pi_i) \leq \frac{M}{B}$ holds, which happens exactly once per $\pi_i$. To capture the time needed to extend the prefixes, we rephrase the internal loop to repeat $\sigma^{j+1}$ times instead of $\sigma^j$. In lines $9-10$ we assume constant time insertions and deletions from $P'$.

In line $12$ the integer sorting algorithm is called. In lines $13$-$22$ the First-Fit Decreasing heuristic for bin packing problem \cite{Yue1991} is implicitly used to construct the virtual trees. Notice that $G$ in the algorithm stands for the content of a bin. Because $\frac{M}{\sigma} < Bf(\pi_i) \leq M$ the external loop is iterated between $\frac{NB}{\sigma M}$ times in the best and $\frac{NB}{M}$ times in the worst case.

\begin{algorithm}[htb]
\KwIn{Input string $S$, alphabet $\Sigma$, $1^{st}$ level memory size $M$}
\KwOut{Set of $VirtualTrees$}
\DontPrintSemicolon
$VirtualTrees \leftarrow \emptyset$

$P \leftarrow \emptyset$

$P' \leftarrow \{ \forall $ symbol $s \in \Sigma$ generate $\pi_i \in P' \}$

\Repeat{$P' = \emptyset$}{
	scan input string $S$
	
	count in $S$ the frequency $f(\pi_i)$ of every $\pi_i \in P'$
	
	\ForAll{$\pi_i \in P'$}{
		\lIf{$0 < Bf(\pi_i) \leq M$} { add $\pi_i$ to $P$ }
			
		\lElse{ \lForAll{symbol $s \in \Sigma$}{ add $\pi_i s$ to $P'$}
		
		remove $\pi_i$ from $P'$
		}
	}
}

sort $P$ in descending $f(\pi_i)$ order

\Repeat{$P = \emptyset$}{
	$G \leftarrow \emptyset$
	
	add $P.head$ to $G$ and remove the item from $P$

	$\pi_{curr} \leftarrow$ next item in $P$

	\While{$NOT$ end of $P$}{
		\If{$f(\pi_{curr}) + SUM_{\gamma \in G}(f(\gamma)) \leq \frac{M}{B}$}{
			add $\pi_{curr}$ to $G$ and remove the item from $P$		
		}
	
		$\pi_{curr} \leftarrow$ next item in $P$ 		
	}	
	add $G$ to $VirtualTrees$
}

\Return{$VirtualTrees$}
\caption{VerticalPartitioning}
\label{alg:verticalpartitioning}
\end{algorithm}

\paragraph{Time complexity}

By exploiting the geometric sum $\sigma + \sigma^2 + ... + \sigma^j = \frac{\sigma^{j+1}-1}{\sigma-1}$ the first loop in lines $4$-$11$ overall takes:

$\sum\limits_{j=1}^{\log_{\sigma} \frac{NB}{M}} \left( Scan(N) + \sigma^{j+1} \right) = \log_{\sigma} \frac{NB}{M} \cdot Scan(N) + \frac{\sigma^2 (NB-M)}{M \cdot \sigma - M},$

that is bounded by $O \left( N \log_{\sigma} \frac{NB}{M} + \frac{\sigma B N}{M} \right)$.

We assume a comparison-based sorting used in line $12$ running in $O(|P| \lg |P|) = O \left( \frac{NB}{M} \lg \frac{NB}{M} \right)$ time. First-Fit Decreasing heuristics in lines $13$-$22$ fits $|P|$ prefixes to bins of size $M$ in time $O\left( \left(\frac{NB}{M} \right)^2 \right)$ in the worst case. Adding all up, the whole vertical partitioning phase requires

\begin{equation}
O \left( N \log_{\sigma} \frac{NB}{M} + \frac{\sigma NB}{M} + \left( \frac{NB}{M} \right)^2 \right)
\label{eq:vertpart-time}
\end{equation}
time in the worst case for the uniformly random input text. 

\paragraph{I/O complexity}

In lines $4$-$11$, the external memory is accessed in line $6$ when reading the input string and requiring $Scan(N)$ I/Os. Intermediate prefix frequencies are stored inside $P'$. Each counter requires $\lceil \lg n \rceil$ bits and on $j^{th}$ iteration of external loop we have an expected number of $|P'|=\sigma^j$ counters present. When updating the counters, two cases are possible on each prefix occurrence: 1) If $M \geq |P'|\lg n$, no cache misses occur for updating the counters. 2) If $M<|P'| \lg n$, then we have $\frac{M}{|P'| \lg n}$ probability of a cache miss. Because $|P'|$ is growing exponentially, the last iteration at $j=\log_{\sigma} \frac{NB}{M}$ and $|P'|=\frac{NB}{M}$ will asymptotically capture cache misses of all the previous iterations. Storing all $N$ occurrences in the last iteration will incur $N \cdot \frac{M}{|P'| \lg n} = N \cdot \frac{M \cdot M}{NB \lg n} = \frac{M^2}{B \lg n}$ accesses to missing elements. Since data is transferred in blocks, $B$ times more elements will be prefetched on each access, so the probability of a block miss becomes $\frac{M^2}{B^2 \lg n}$. In lines $7$-$10$, no cache misses occur, if $|P'| \leq M$ and for $|P'|>M$, a single scan over $P'$ is done to check the frequencies and correspondingly update the data structure. Asymptotically this term is overtaken by $Scan(N)$ operation in line $6$. Overall, this leads to the following I/O complexity of lines $4$-$11$:

\begin{equation*}
\begin{split}
&M \geq \sqrt{NB \lg N}: O \left( \log_{\sigma} \frac{NB}{M} \cdot \frac{N}{B} \right) \\
&M<\sqrt{NB \lg N}: O \left( \log_{\sigma} \frac{NB}{M} \cdot \left( \frac{N}{B} + \frac{M^2}{B^2\lg n} \right) \right) = O \left( \log_{\sigma} \frac{NB}{M} \cdot \frac{N}{B} \right)
\end{split}
\end{equation*}

The sorting in line $12$ is done in-memory, if $M \geq \sqrt{NB \lg N}$ and does not require any I/Os. Otherwise, one of the I/O efficient external sorting algorithms is used for sorting $|P|=\frac{NB}{M}$ elements requiring $O(\frac{N \cdot B}{M \cdot B} \log_{\frac{M}{B}} \frac{N \cdot B}{M \cdot B}) = O(\frac{N}{M} \log_{\frac{M}{B}} \frac{N}{M})$ I/Os.

The heuristic for virtual tree construction accesses $P$ and a virtual tree $G$, $|G| \leq \frac{M}{B}$. Following the Assumption \ref{assumption:random-text}, we expect $|G|=\left[1, \sigma\right)$. Since each prefix size is expected to be $\lg N$ bits, then the expected size of $G$ is $\sigma \lg N$ bits in the worst case which may or may not fit $M$. In any case, all the $\frac{NB}{M}$ prefixes from $P$ of size $\frac{NB \lg N}{M}$ need to be stored to $G$ eventually, and additions can be made sequential which leads to $\frac{N\lg N}{M}$ overall expected block transfers for constructing all $G$ instances. If $|P| \lg N \leq M$, then no cache misses occur in lines $13$-$22$. Otherwise, $P$ is always accessed in the sequential order in lines $17$-$20$ requiring $\frac{|P| \lg N}{B}=\frac{N \lg N}{M}$ I/Os. Since external loop in lines $13$-$22$ is executed $O(|P|)$ times in the worst case, the quadratic $P$'s I/O complexity overtakes the $G$'s one.

The whole vertical partitioning has an expected I/O complexity:
\begin{equation}
\begin{split}
&M \geq \sqrt{NB \lg N}: O \left( \frac{N}{B} \log_{\sigma} \frac{NB}{M} \right)  \\
&M<\sqrt{NB \lg N}: O \left( \frac{N}{B} \log_{\sigma} \frac{NB}{M} + \frac{N}{M} \log_{\frac{M}{B}} \frac{N}{M} + \frac{NB^2 \lg N}{M^2}\right)
\end{split}
\label{eq:verpart-io}
\end{equation}

\subsection{Horizontal Partitioning}

Algorithm \ref{alg:horizontalpartitioning} shows the horizontal partitioning algorithm taken from \cite[pp.\ $8$]{Mansour2011}. The relative suffix array $SA$ maps the suffix position in lexicographically ordered list of all suffixes in a suffix subtree corresponding to $\pi$ to the position in the input string. $ISA$ denotes the inverse suffix array defined as $ISA[ SA[i] ] = i$. The relative longest common prefix array $LCP$ contains the longest common prefix length of consecutive suffixes in $SA$. We denote by $n \leq \frac{M}{B}$ the number of suffixes with unfinished branches: the final path from the root of the suffix subtree to the leaf corresponding to the suffix was not constructed yet. In general, $n$ decreases over external loop iterations. However, by Assumption \ref{assumption:random-text} $n$ remains at the initial value all the time until the final iteration. The array $A$ contains additional information on $n$. If $A[i]=A[i+1]$, suffixes corresponding to index $i$ and $i+1$ belong to the same path from the root to the node down to the current depth.

\begin{algorithm}[htb]
\KwIn{Input string $S$, prefix $\pi$}
\KwOut{Arrays $SA$ and $LCP$ corresponding suffix sub-tree $\mathcal{T}_\pi$}
\DontPrintSemicolon

$SA$ contains the locations of prefix $\pi$ in string $S$

$LCP \leftarrow \{\}$

$ISA \leftarrow \{0,1,...,|SA|-1\}$

$A \leftarrow \{0,0,...,0\}$

$Buf \leftarrow \{\}$

$P \leftarrow \{0,1,...,|SA|-1\}$

$start \leftarrow |\pi|$

\While{there exists an undefined $LCP[i]$, $1 \leq i \leq |SA|-1$}{
	$range \leftarrow GetRangeOfSymbols$
	
	\For{$i \leftarrow 0$ to $|SA|-1$}{
		\If{$ISA[i] \neq done$}{
			$Buf[ ISA[i] ] \leftarrow ReadRange(S, SA[ ISA[i] ]+start, range)$
			\newline // ReadRange(S,a,b) reads $b$ symbols of $S$ starting at position $a$
		}	
	}
	\For{every active area $AA$}{
		Reorder the elements of $Buf$, $P$ and $SA$ in $AA$ so that $Buf$ is lexicographically sorted. In the process maintain the index $ISA$
		
		If two or more elements $\{a_1,...,a_t\} \in AA, 2 \leq t$, exist such that $Buf[a_1]=...=Buf[a_i]$ introduce for them a new active area	
	}
	\For{all $i$ such that $LCP[i]$ is not defined, $1 \leq i \leq |SA|-1$}{
		$cp$ is the common prefix of $Buf[i-1]$ and $Buf[i]$
		
		\If{$|cp|<range$}{
			$LCP[i] \leftarrow (Buf[i-1][|cp|], Buf[i][|cp|], start+|cp|)$
			
			\If{$LCP[i-1]$ is defined or $i=1$}{
				Mark $ISA[ P[i-1]]$ and $A[i-1]$ as $done$
			}
			\If{$LCP[i+1]$ is defined or $i=[SA]-1$}{
				Mark $ISA[ P[i] ]$ and $A[i]$ as $done$ // last element of an active area
			}
		}
	}
	$start \leftarrow start + range$
}

\Return{($SA$,$LCP$)}

\caption{HorizontalPartitioning.SubTreePrepare}
\label{alg:horizontalpartitioning}
\end{algorithm}

In line $9$ the algorithm determines the string chunks length $range$, such that $B \leq range \leq \frac{M}{n}$. In lines $10$-$12$ the algorithm reads chunks from the input string at positions corresponding to the ends of unfinished branches and stores them into $Buf$. In lines $13$-$15$ an in-memory string sorting of $Buf$ is done constructing $SA$ and $ISA$. Finally in lines $16$-$23$ the in-memory construction of $LCP$ is done by calculating the common prefix $cp$ of $SA[i]$ and $SA[i+1]$ for all $i$. The external while loop at lines $8$-$24$ is repeated depending on the similarity of the string chunks inside $Buf$. Under Assumption \ref{assumption:random-text}, the expected longest common prefix length of $n$ uniformly random strings is $\lceil \log_\sigma n \rceil$. This is also the amount of characters we need to read for each string in order to discriminate between them. Finally, by reading $range \geq B$ characters per each suffix, the total expected number of the external while loop iterations is bounded by $O(\frac{1}{B} \log_{\sigma} \frac{M}{B})$.

\paragraph{Time complexity}

In line $9$ we require $O(1)$ time to calculate $range$. In lines $10$-$12$ we require $range \cdot n$ time to fill $n$ buffers assuming reading each character requires constant time. String sorting in lines $13$-$15$ takes $O(n \cdot D)$ time, where $D$ denotes the \emph{distinguishing prefix size} (see for example \cite{Bingmann2013}). Under Assumption \ref{assumption:random-text}, $D=\log_\sigma n$, so we need $O(n \log_\sigma n)$ time to sort $n$ strings each iteration. Lines $16$-$23$ take $O(n \cdot range)$ time in the worst case. Under Assumption \ref{assumption:random-text} it is only used once in the last iteration of the external while loop and it takes $O(n \log_\sigma n) = O(M)$ time which is overtaken by the previous term.

The horizontal partitioning algorithm is embarrassingly parallel, so a $p$-fold speedup is expected using $p$ processors. Each processor dispatches a new process for the assigned virtual tree. Taking all $O\left( \frac{NB}{M} \right)$ virtual trees into account, $O\left( \frac{1}{B} \log_{\sigma} \frac{M}{B} \right)$ iterations of the external loop, and $n=\frac{M}{B}$, the expected execution time to construct all suffix subtrees in parallel is

\begin{equation}
O \left( \frac{1}{p} \frac{NB}{M} \frac{1}{B} \log_\sigma \frac{M}{B} \left( M + \frac{M}{B} \log_{\sigma} \frac{M}{B} \right) \right) = O \left( \frac{N}{p} + \frac{N}{Bp} \log_{\sigma}^2 \frac{M}{B} \right)
\label{eq:horizontalpart-time}
\end{equation}

\paragraph{I/O complexity}

Determining $range$ in line $9$ requires no I/Os. I/O complexity of lines $10$-$12$ depends on $N$ in relation to $M$ and $B$. Under Assumption \ref{assumption:random-text}, $n=\frac{M}{B}<\frac{N}{B}$, then all $n$ occurrences are at least $B$ bytes apart on average in the input string and we require $n$ block transfers to read all string chunks. In lines $13$-$23$ no external memory accesses are required since all $SA$, $ISA$, $LCP$ and $Buf$ are, by definition, small enough to fit $M$. The expected I/O complexity of a sequential Algorithm \ref{alg:horizontalpartitioning} is $O\left( \frac{M}{B} \cdot \frac{1}{B} \log_{\sigma} \frac{M}{B} \right) = O(\frac{M}{B^2} \log_\sigma \frac{M}{B})$ I/Os.

The final step of the horizontal partitioning omitted in this paper is the in-memory suffix subtree construction from the obtained $SA$ and $LCP$ arrays and writing it to disk by a depth-first tree traversal \cite[pp.\ 8]{Mansour2011}. It requires linear time for the construction and a single $Scan(M)$ I/Os for storing it.

The horizontal partitioning phase is embarrassingly parallel also from the input string access point of view on a local machine since the file system and I/O schedulers allow the device to serve multiple requests using a single scan. The Parallel I/O complexity for constructing $O\left(\frac{NB}{M}\right)$ virtual trees is then:

\begin{equation}
O\left( \frac{1}{p} \frac{NB}{M} \frac{M}{B^2} \log_{\sigma} \frac{M}{B} \right) = O\left( \frac{N}{Bp} \log_{\sigma} \frac{M}{B} \right)
\label{eq:horizontalpart-io}
\end{equation}

\section{Empirical evaluation}
\label{chap:empirical_evaluation}

All our experiments involved measuring wall-clock running times of specific parts of ERA taking a uniformly random string as an input. The experiments were executed on a single machine with two 16-core AMD Opteron 6272 processors at $2.4$ GHz and $128$ GiB RAM. The experiments were run in parallel using Open MPI library version 1.4.6. Input data were read from and the results were stored to a $4$ TB Hitachi hard drive model HGST HDN724040ALE640 where we flushed all the I/O caches each time a suffix tree was written to a file assuring no parts of the file remained in the system RAM. Unfortunately we could only use the input sets of sizes up to $3.5$ GiB since the original ERA algorithm uses $32$-bit pointers and cannot address larger input strings.

\begin{figure}
\begin{subfigure}{.5\textwidth}
  \centering
  \includegraphics[width=0.95\linewidth]{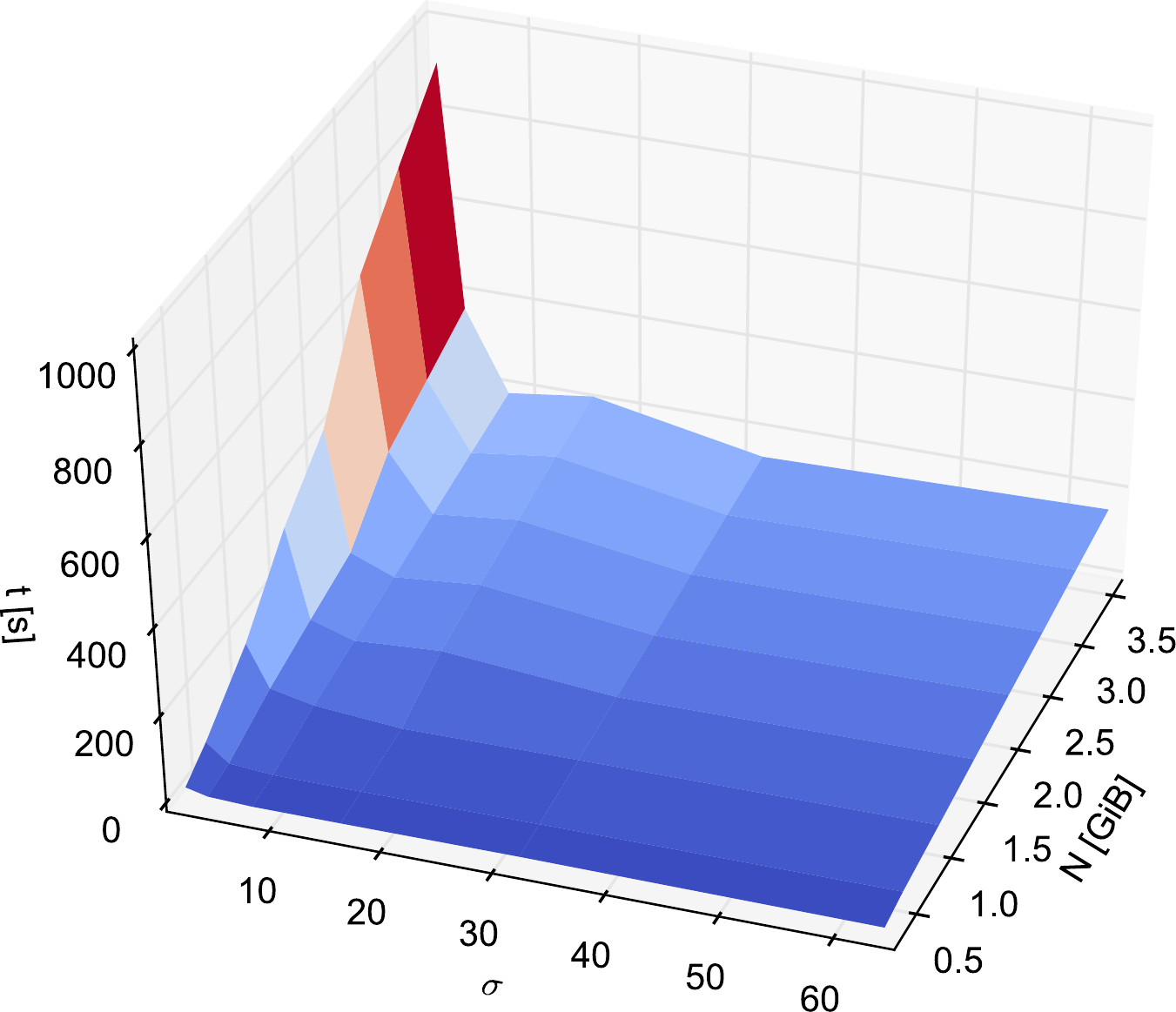}
  \caption{Vertical partitioning}
  \label{fig:bm1-vertpart}
\end{subfigure}%
\begin{subfigure}{.5\textwidth}
  \centering
  \includegraphics[width=0.95\linewidth]{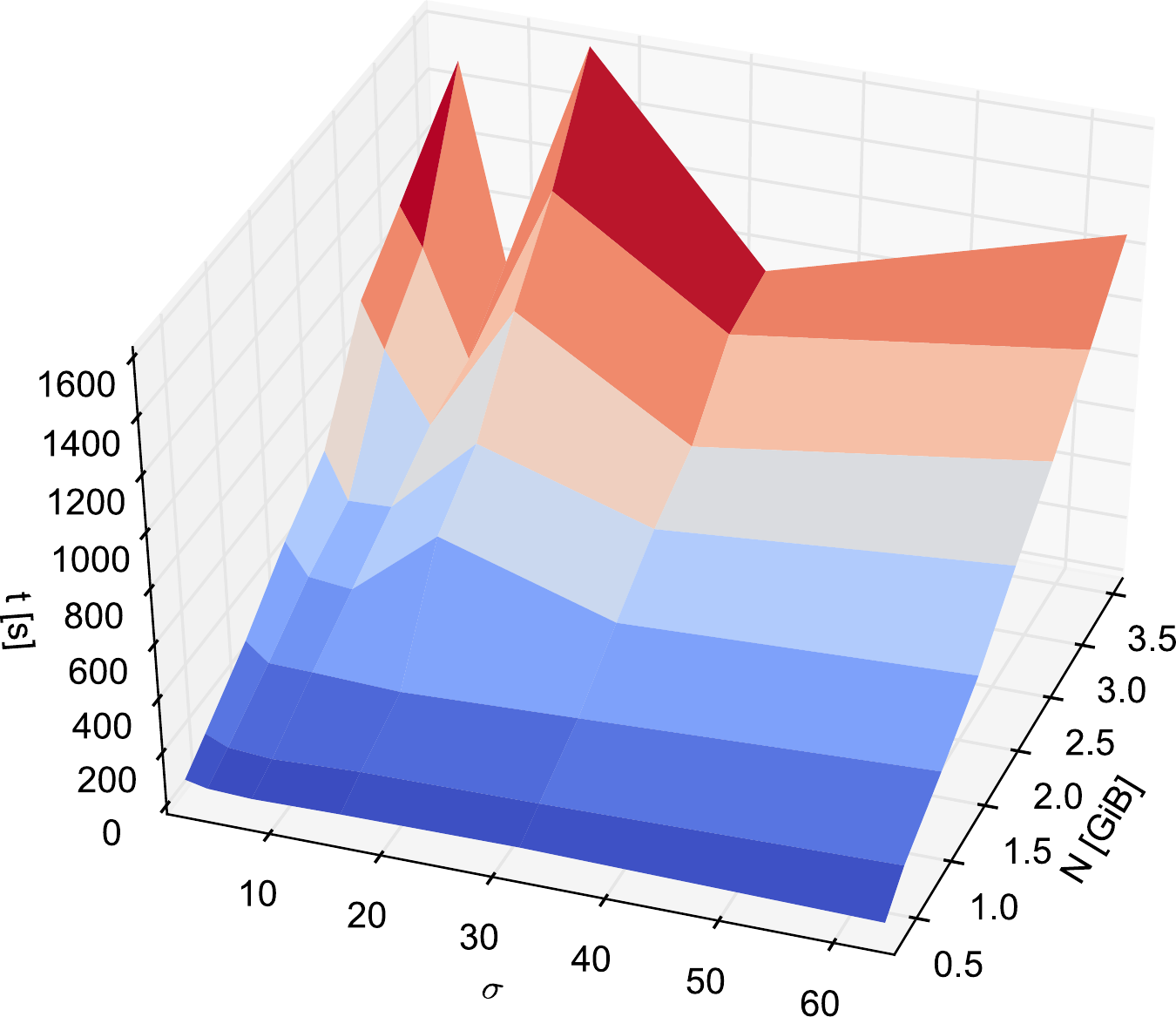}
  \caption{Horizontal partitioning}
  \label{fig:bm1-horizontalpart}
\end{subfigure}
\caption{Running times in seconds of the vertical and horizontal partitioning phases for $p=32$, $N=[256 \text{MiB}, ..., 3.5 \text{GiB}]$, $M=512 \text{MiB}$, and $\sigma=[2,...,64]$.}
\label{fig:bm1}
\end{figure}

Figure \ref{fig:bm1} shows the execution times of ERA for both the vertical and the horizontal partitioning. The vertical partitioning running times grow almost linearly with the size of the input $N$. By increasing the alphabet size the execution times decreases. This behaviour fits the Equations \ref{eq:vertpart-time} and \ref{eq:verpart-io}.

\begin{figure}[htb]
  \begin{center}
    \leavevmode
    \includegraphics[width=\textwidth]{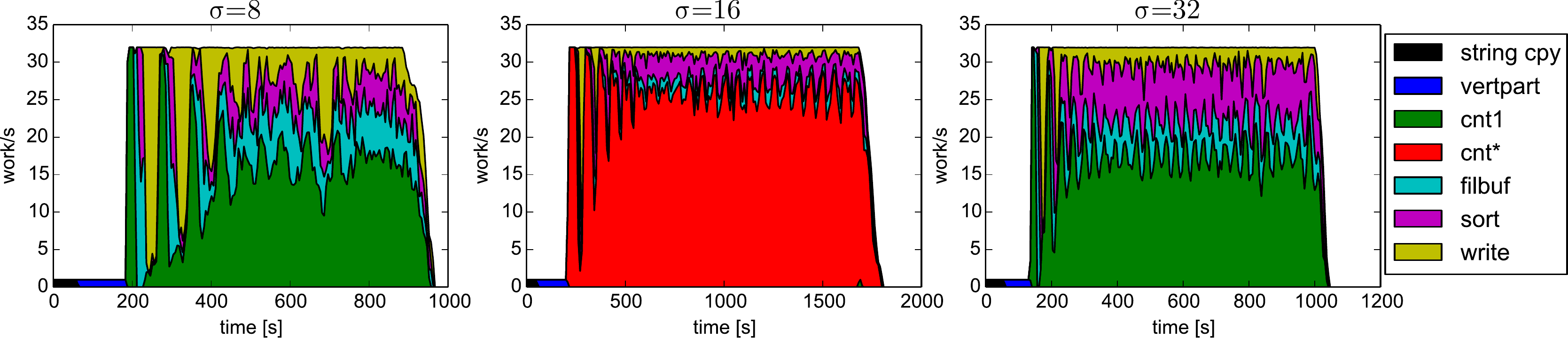}
  \end{center}
  \caption{Amount and type of work invested for each phase of ERA during the execution for $p=32$, $N=3584 \text{MiB}$, $M=512 \text{MiB}$, and $\sigma=\{8, 16, 32\}$.}
  \label{fig:cputimes}
\end{figure}

The horizontal partitioning behaviour is more surprising. The execution time increases almost linearly for increasing $N$ which fits Equations \ref{eq:horizontalpart-time} and \ref{eq:horizontalpart-io}. By increasing the alphabet size however, the execution time doesn't necessarily decrease. The worst case execution time is for $\sigma=16$. Figure \ref{fig:cputimes} shows the amount and type of work done during ERA execution. The \texttt{cnt1} and \texttt{cnt*} work corresponds to line $1$ in Algorithm \ref{alg:horizontalpartitioning} for the vertical tree of size $1$ or $>1$ respectively. The complexity of \texttt{cnt*} code is obviously greater than \texttt{cnt1}. In our analysis however, we assumed the line $1$ to require a single scan over the input string.

\section{Discussion and Conclusion}
\label{chap:discussion_and_conclusion}

We showed that for uniformly random input strings ERA execution time is bounded by $O \left( N \log_{\sigma} \frac{NB}{M} + \frac{\sigma NB}{M} + \left( \frac{NB}{M} \right)^2 \right)$ time and $O \left( \frac{N}{B} \log_{\sigma} \frac{NB}{M} \right)$ I/Os for the vertical partitioning, and $O \left( \frac{N}{p} + \frac{N}{Bp} \log_{\sigma}^2 \frac{M}{B} \right)$ time and $O\left( \frac{N}{Bp} \log_{\sigma} \frac{M}{B} \right)$ I/Os for the horizontal partitioning phase. First, according to \cite[pp.\ 5]{Mansour2011} and a private discussion with ERA authors, the preferred size of $M$ should be an order of a few hundred mega bytes in order to achieve good performance. We formally showed in Equation \ref{eq:verpart-io} the I/O complexity of the vertical partitioning increases significantly for $M < \sqrt{B N \lg N}$ which explains the authors' dilemma.

Secondly, the ERA's time and I/O complexity even for random input strings is larger than the sorting lower bound in two ways: 1) the vertical partitioning is done sequentially only, and 2) focusing on I/O complexity, both the vertical and the horizontal partitioning phases require an additional logarithmic factor $O(\log_\sigma \frac{NB}{M})$ and $O \left( \log_\sigma \frac{M}{B} \right)$ respectively more I/Os than the sorting lower bound. We predict the issue 1) could be solved by splitting the input string to $p$ chunks, each processor reading the frequency of the prefixes in its own chunk in parallel and finally doing a sum of all the frequencies in $\frac{NB}{M} \lg p$ steps. Solving the issue 2) is more demanding. The vertical partitioning factor originates from extending the prefixes only by a single character at a time. This might be solved by analysing the text first in an ``intelligent way'' and immediately picking the ideal prefix length. The additional logarithmic factor in the horizontal partitioning originates from the fact that string sorting is slower than the suffix sorting, because all the existing work being done for sorting suffixes in the past is ignored for the suffixes to come. One can design an optimal suffix tree or suffix array construction algorithm by employing an efficient suffix sorting only, for example employing \emph{the induced sorting} principle as used in eSAIS algorithm.

Finally, we should discuss input strings other than the random ones. Let the skewness of the text be defined as the length of the \emph{longest repeated substring} (LRS) in the string. The expected $|LRS|=\log_\sigma N$ characters in a uniformly random string of length $N$. If we concatenate two or more human genomes, $|LRS|$ effectively becomes $O(N)$. The running time of ERA and also of WF algorithm becomes quadratic since string prefixes in the horizontal partitioning phase will not be unique until the final delimiter character is reached. Even more, the PCF and B$^2$ST algorithm suffer as well since the required partitioning with unique partition beginnings cannot be done. eSAIS suffix array construction algorithm, which is within the time and I/O sorting bounds in the worst case, handles the case fine \cite[see the ``skyline'' input on pp.\ 10]{Bingmann2013-2}. The quest for designing a parallel algorithm for the suffix tree or suffix array construction which is theoretically within the suffix sorting bounds and performs fast in practice is thus still open.

\bibliographystyle{splncs03}
\bibliography{jekovec}

\begin{thebibliography}{10}
\providecommand{\url}[1]{\texttt{#1}}
\providecommand{\urlprefix}{URL }

\bibitem{Abouelhoda2004}
Abouelhoda, M.I., Kurtz, S., Ohlebusch, E.: {Replacing suffix trees with
  enhanced suffix arrays}. Journal of Discrete Algorithms  2(1),  53--86 (2004)

\bibitem{Aggarwal1988}
Aggarwal, A., {Vitter, Jeffrey}, S.: {The input/output complexity of sorting
  and related problems}. Communications of the ACM  31(9),  1116--1127 (1988)

\bibitem{Arge2008}
Arge, L., Goodrich, M.T., Nelson, M., Sitchinava, N.: {Fundamental parallel
  algorithms for private-cache chip multiprocessors}. In: Proceedings of the
  20th annual symposium on Parallelism in algorithms and architectures - SPAA
  '08. p. 197. ACM Press, New York, USA (2008)

\bibitem{Barsky2009}
Barsky, M., Stege, U., Thomo, A., Upton, C.: {Suffix trees for very large
  genomic sequences}. In: Proceeding of the 18th ACM conference on Information
  and knowledge management - CIKM '09. p. 1417. ACM Press, New York, USA (2009)

\bibitem{Bingmann2013-2}
Bingmann, T., Fischer, J., Osipov, V.: {Inducing Suffix and LCP Arrays in
  External Memory}. In: Proceedings of the 15th Meeting on Algorithm
  Engineering and Experiments, ALENEX 2013. pp. 88--102. Society for Industrial
  and Applied Mathematics (2013)

\bibitem{Bingmann2013}
Bingmann, T., Sanders, P.: {Parallel String Sample Sort}. Algorithms - ESA 2013
  pp. 169--180 (2013)

\bibitem{Comin2013}
Comin, M., Farreras, M.: {Efficient parallel construction of suffix trees for
  genomes larger than main memory}. In: Proceedings of the 20th European MPI
  Users' Group Meeting on - EuroMPI '13. p. 211. ACM Press, New York, New York,
  USA (2013)

\bibitem{Farach-Colton2000}
Farach-Colton, M., Ferragina, P., Muthukrishnan, S.: {On the sorting-complexity
  of suffix tree construction}. Journal of the ACM  47(6),  987--1011 (nov
  2000)

\bibitem{Ghoting2009}
Ghoting, A., Makarychev, K.: {Serial and parallel methods for I/O efficient
  suffix tree construction}. In: Proceedings of the 35th SIGMOD international
  conference on Management of data - SIGMOD '09. p. 827. ACM Press, New York,
  USA (2009)

\bibitem{Heinz2002}
Heinz, S., Zobel, J., Williams, H.E.: {Burst tries: a fast, efficient data
  structure for string keys}. ACM Transactions on Information Systems  20(2),
  192--223 (apr 2002)

\bibitem{Manber1990}
Manber, U., Myers, G.: {Suffix arrays: a new method for on-line string
  searches}. In: Proceedings of the first annual ACM-SIAM symposium on Discrete
  algorithms. pp. 319--327. Society for Industrial and Applied Mathematics (jan
  1990)

\bibitem{Mansour2011}
Mansour, E., Allam, A., Skiadopoulos, S., Kalnis, P.: {ERA: efficient serial
  and parallel suffix tree construction for very long strings}. Proceedings of
  the VLDB Endowment  5(1),  49--60 (2011)

\bibitem{Szpankowski1993}
Szpankowski, W.: {A generalized suffix tree and its (un)expected asymptotic
  behaviors}. SIAM Journal on Computing  22(6),  1176--1198 (1993)

\bibitem{Vitter1994}
Vitter, J.S., Shriver, E.: {Algorithms for parallel memory, I: Two-level
  memories}. Algorithmica  12(2-3),  110--147 (1994)

\bibitem{Weiner1973}
Weiner, P.: {Linear pattern matching algorithms}. In: Switching and Automata
  Theory, 1973. SWAT'08. IEEE Conference Record of 14th Annual Symposium on.
  pp. 1--11. IEEE (1973)

\bibitem{Yue1991}
Yue, M.: {A simple proof of the inequality FFD (L) ≤ 11/9 OPT (L) + 1, ∀L
  for the FFD bin-packing algorithm}. Acta Mathematicae Applicatae Sinica
  7(4),  321--331 (1991)

\end{thebibliography}

\end{document}